%% file: main.tex
\documentclass[11pt]{article}
\usepackage[margin=1in]{geometry}
\usepackage{authblk} 
\usepackage{setspace} 
\usepackage{cite}
\usepackage{amsthm}
\usepackage{amsmath,amssymb,amsfonts}
\usepackage{algorithmic}
\usepackage{textcomp}
\usepackage{graphics} 
\usepackage{mathptmx} 
\usepackage{times} 
\usepackage{algorithmic}
\usepackage{textcomp}
\usepackage[english]{babel}
\usepackage{grffile}
\usepackage[justification=centering]{caption}
\newtheorem{theorem}{Theorem}
\newtheorem{lemma}{Lemma}

\usepackage[normalem]{ulem}
\title{Prescribed-Time Event-Triggered Control for Matrix-Scaled Networks}
\author[1]{K. P. Sunny }
\author[2]{Rakesh R. Warier}

\affil[1]{Department of Electrical Engineering, NIT Calicut, Kerala, India\\ \texttt{sunny\_p240252ee@nitc.ac.in}}
\affil[2]{Faculty of Electrical Engineering,  NIT Calicut, Kerala, India\\ \texttt{rakeshw@nitc.ac.in}}

\begin{document}
\date{}
\maketitle
\begin{abstract}
This article proposes a distributed control method for matrix-scaled multi-agent networks aimed at achieving convergence within a user-defined time frame. The control law of each individual agent relies only on information from neighboring agents and is updated at discrete intervals determined by state-dependent triggering functions, reducing the frequency of agent interactions. To this end, first, the controller is augmented with a time-varying gain. Then, the dynamics of the closed-loop system over the finite-time interval is transformed into an infinite-time frame using time scaling. Lyapunov-based analysis is employed to derive suitable triggering conditions that guarantee the asymptotic convergence of the time-transformed system, thereby ensuring the prescribed-time convergence of the original system.  
\end{abstract}

\section{Introduction}
\label{sec:introduction}
Coordinated behaviors such as flocking in birds, schooling in fish, colony formation in social insects, etc., where each entity within a group acts collectively without central control, are frequently observed in nature.  These natural phenomena have inspired researchers to develop different control policies for networked systems to achieve various coordination objectives such as consensus, leader following, and  formation control (see \cite{survey_general} and references therein, \cite{secure}). 

Consensus in multi-agent systems, where agents converge to a common state, has been extensively studied in control theory. However, many engineering applications—such as multi-scaled space vehicle coordination \cite{zhao2014similitude}, compartmental systems\cite{haddad2010nonnegative}, and distributed resource allocation — often require agents to reach predefined scaled ratios rather than exact agreement. To address this need, scaled consensus was introduced \cite{ROY2015259}, where agents converge to states aligned along a straight line with specified proportions. This concept has been applied in multi-vehicle coordination \cite{liu2025} and hierarchical control of microgrid \cite{mo2022hierarchical}. Generalizing this framework, matrix-scaled consensus (MSC) was introduced in \cite{trinh_msc_conf}, which allows each agent to have a scaling matrix, enabling convergence to steady states defined by these matrices. MSC unifies various consensus paradigms— including standard consensus, bipartite consensus, and multi-cluster consensus—through appropriate matrix selection without altering the structure of the control law \cite{trinh_msc}. MSC can also be employed to model multi-dimensional opinion dynamics where the scaling matrices represent the private belief system of an individual on logically dependent topics \cite{trinh_msc_conf}. Moreover, matrix-scaled consensus permits agents to form multiple clusters in prespecified geometric configurations, making it valuable in applications such as multi-team robotic encirclement, containment control \cite{zhang2024containment}, and dynamic formation reconfiguration. Existing works on MSC include adaptive control \cite{trinh_msc}, sliding mode control \cite{MSCuncertainnetworks}, model reference adaptive control \cite{secondorderdirected} and state constrained distributed control \cite{MSCWeightedNetworksWithStateConstraintsstateconstraints} among others.

However, in many cases like rendezvous, cooperative engagement, formation switching, etc., the convergence must occur at a specific, user-defined time. As an attempt, finite-time \cite{finitetimehomogenous}, and fixed-time \cite{fixedtimeoverview} control methods are applied to multi-agent systems. They render non-asymptotic convergence, but with finite-time control, the settling time depends on initial conditions which cannot be estimated in practice, whereas fixed-time controllers provide only an upper bound which is often overestimated and not easily tunable. Recently, the notion of prescribed-time control introduced in \cite{PTCfirst}, has been extended to solve consensus objectives in networked systems \cite{PTcontrainmentcontrol, PTcooperativeengagement, ptcreviewlewis,su2025prescribed, decentralized2025}. The method allows the user to preset equilibration time, independent of system parameters and initial conditions. In addition, the control law is smooth and robust to external disturbances and nonvanishing perturbations.

While control over settling time is desirable, it is equally important in networked systems to reduce computational and communication overhead by minimizing local agent interactions. Hence, event-triggered control strategies have been developed to reduce network utilization and computational burden on individual agents \cite{eventtrigger, eventtriggerold}. The authors of \cite{timetrnasformYucelen} adopted the time scaling method to design a prescribed-time event-triggered controller for tracking  time varying commands. Although the triggering threshold is flexible, incessant communication is observed near the terminal time. An event-based prescribed-time controller is developed for cluster consensus in \cite{ptceventcluster}. When the dynamics contain uncertainties and time delays, dynamic event-triggering is more suitable to achieve practical prescribed-time tracking consensus \cite{ptcdynamicevent}. The authors of \cite{eventmatrixweighted} applied event-triggering technique to reduce agent interactions in matrix-weighted networks, a topology in which the edges of the network are assigned matrix-valued weights.  

 The remainder of this article is organized as follows. Section 2 introduces the basic notation and the problem statement. The proposed control policy and implementation aspects are presented in Section 3. Section 4 concludes the article.

\section{Problem Statement and Preliminaries}
\textit{Notations:}
Throughout this article, $\mathbb{R}$, $\mathbb{R}^+$, $\mathbb{Z}_{0}$, and $\mathbb{N}$ denote the sets of real numbers, positive real numbers, non-negative integers, and natural numbers, respectively. We use $\mathbb{R}^n$ and $\mathbb{R}^{m \times n}$ to represent $n$-dimensional vectors and $m \times n$ real matrices. Vectors of all ones and zeros are denoted by $1_n$ and $0_n$, respectively. The transpose, kernel, and image of a matrix $A$ are given by $A^T$, $\ker(A)$, and $\operatorname{im}(A)$. The Euclidean norm is denoted by $\| \cdot \|$, and the smallest and largest eigenvalues of a matrix are $\lambda_{\min}(\cdot)$ and $\lambda_{\max}(\cdot)$. A matrix $A \in \mathbb{R}^{n \times n}$ is positive (resp., negative) definite if $x^T A x > 0$ (resp., $< 0$) for all nonzero $x \in \mathbb{R}^n$; the signum function $\operatorname{sgn}(A)$ returns $1$ if $A$ is positive definite and $-1$ if negative definite. The Kronecker product is denoted by $\otimes$.

\textit{Network graph:} The inter-agent topology is modeled by a connected, undirected graph $\mathcal{G} = (\mathcal{V}, \mathcal{E}, \mathcal{A})$, where $\mathcal{V} = {1, \dots, n}$ is the node set, $\mathcal{E} \subseteq \mathcal{V} \times \mathcal{V}$ the edge set, and $\mathcal{A} = [a_{ij}]$ the adjacency matrix with $a_{ij} = 1$ if $(i,j) \in \mathcal{E}$, and $0$ otherwise. The graph is symmetric ($a_{ij} = a_{ji}$) and loop-free ($a_{ii} = 0$). The neighbor set of node $i$ is $\mathcal{N}_i = \{ j \in \mathcal{V} \mid (i,j) \in \mathcal{E} \}$, with cardinality $|\mathcal{N}_i|$. The degree matrix $\mathcal{D} = \operatorname{diag}(|\mathcal{N}_1|, \dots, |\mathcal{N}_n|)$ and the Laplacian $\mathcal{L} = \mathcal{D} - \mathcal{A}$ satisfy $0 = \lambda_1 < \lambda_2 \leq \cdots \leq \lambda_n$, where $\ker(\mathcal{L}) = \operatorname{im}(1_n)$.
\begin{lemma}\label{lemma1} [Young's inequality]
      Let $x,y\in \mathbb{R}^{n}$, then, for any scalar $a>0$, one has $ x^{T}y\le(x^{T}x)/(2a)+(a/2)y^{T}y$.
\end{lemma}

\textit{Problem Statement:}
In this article, we consider a networked system of $n\in \mathbb{N}$ agents interacting over an undirected and connected graph $\mathcal{G}$. The dynamics of each agent $i\in \mathcal{V}$ is governed by a $d$-dimensional state vector $x_{i} \in \mathbb{R}^{d}$ and a positive or negative definite scaling matrix $S_{i}\in \mathbb{R}^{d \times d}$ \cite{trinh_msc_conf}. Let $|S_{i}|\equiv sgn(S_{i})S_{i}$, which is positive definite. Further, one has $sgn(S_{i})=sgn(S_{i}^{T})=sgn(S_{i}^{-1})$, and $|S_{i}^{-1}|=|S_{i}|^{-1}$. We define $\text{x}=\begin{bmatrix}
  x_{1}^{T} & x_2^{T} &... & x_{n}^{T} \end{bmatrix}^{T} \in \mathbb{R}^{nd}$, to represent the state vector of the overall system consisting of $n$ agents. Let $sgn(S) = diag\left(\begin{bmatrix}
   sgn(S_{1}) & sgn(S_{2}) & ... & sgn(S_{n})
\end{bmatrix}\right)$, and $\Bar{L}=\mathcal{L}\otimes I_{d}$. We also define $S=blkdiag(S_{1}, S_{2},  .., S_{n})$ and $|S|=blkdiag(|S_{1}|, |S_{2}|,  .., |S_{n}|)$ to represent the respective block diagonal matrices. The network is said to attain matrix-scaled consensus, if and only if there exists a virtual consensus point $x^{v}\in \mathbb{R}^{d}$ and a consensus space $C_s$ defined as
  \begin{align}\label{eq1}
      C_{s}=\{\text{x}\in \mathbb{R}^{dn} | S_{1}x_{1} =S_{2}x_{2} =...=S_{n}x_{n} =x^{v}  \}
  \end{align}
such that $\text{x}(t)$ approaches $C_{s}$ as $t\rightarrow \infty$.
Alternatively, consensus is guaranteed if $\lim_{t \to \infty} (S_{i}x_{i}-S_{j}x_{j}) = 0_{d}$, for all $i,j = 1, 2, .., n$. We consider first-order agents with dynamics
\begin{align}\label{eq02}
    \Dot{x}_{i} = u_{i}, \quad i =1,2,..., n 
\end{align}
where, $x_{i}\in \mathbb{R}^{d}$ and $u_{i} \in \mathbb{R}^{d}$. The standard MSC control protocol \cite{trinh_msc_conf} 
\begin{align}\label{eq403}
    u_{i}=-sgn(S_{i})\sum_{j\in \mathcal{N}_{i}}(S_{i}x_{i}-S_{j}x_{j}),  \quad i=1,2,...n.
\end{align}
renders asymptotic convergence of $\text{x}(t)$ to the set $C_{s}$ with $x^{v}= (\sum_{i=1}^{n}|S_{i}|^{-1})^{-1}\sum_{i=1}^{n}sgn(S_{i})x_{i}(0)$. The equilibrium state of each agent differs from $x^{v}$ by the inverse of its scaling matrix as $lim_{t\rightarrow \infty}x_{i}(t)=S_{i}^{-1}x^{v}$. Due to space constraints, a detailed discussion on the design of the scaling matrices is omitted; further information can be found in \cite{trinh_msc}. 

The control law (\ref{eq403}) guarantees only asymptotic convergence and requires continuous communication among neighboring agents. We focus on designing a prescribed-time event-triggered control strategy for system~(\ref{eq02}) that guarantees convergence within a user-defined time \(T\), reduces inter-agent communication, and ensures Zeno-free behavior. 
 
\section{Distributed prescribed-time event-triggered Control}
In this section, we develop prescribed-time event-triggered control strategies for matrix-scaled networks and establish the stability of the closed-loop system through rigorous mathematical analysis. To this end, we first apply a time transformation $\tau =a(t)$ to transform the dynamics from the finite-time domain $t\in [0,T)$ to the infinite-time domain $\tau \in[0,\infty)$. Let $\alpha (\tau)= a'(a^{-1}(\tau))$. Then, an arbitrary signal $x(t)$ and its derivative $\dot{x}(t)$ in the time frame $t\in [0, T)$ can be expressed in terms of their equivalents in the new time frame $\tau \in [0, \infty)$ as
\begin{equation}
 \begin{aligned}\label{800}
   x(\tau) = & \ x(t) \\
   x'(\tau) = & \frac{dx(\tau)}{d\tau} =  \frac{dx(t)}{dt}\frac{dt}{d\tau} = \frac{1}{\alpha(\tau)}\dot{x}(t) 
\end{aligned} 
\end{equation}
where differentiation with respect to $\tau$ is denoted by $x'$, to distinguish it from the derivatives with respect to $t$ represented by $\dot{x}$. This transformation enables Lyapunov-based stability analysis on an infinite-time horizon, with conclusions extended to the original system \cite{PTcooperativeengagement}.

While prescribed-time controllers permit the user to decide the settling time of the system in advance, it is also often desired to reduce inter-agent communication and computation. This can be achieved if each agent updates its control law only at discrete time instants based on the information about its own state and that of its neighbors. At these instants, the agent also transmits its current state to its neighbors.  

Let the last transmitted state of the agent $i\in \mathcal{V}$ be denoted by $\Hat{x}_{i}(t)$. We define the state measurement error $e_{i}(t)$ as
\begin{align}\label{eq405}
    e_{i}(t) = \Hat{x}_{i}(t)-x_{i}(t)
\end{align}

Let the sequence $\{t_{k}\}= \{0,t_{1}, t_{2}, ... \}$ represent the instances at which the agent samples its state $x_{i}(t)$, computes the measurement error, and evaluates the event-triggering function $f_{i}(e_{i}(t_{k}),\Hat{x}_{j\in N_{i}}(t_{k}))$, for deciding whether to update its own control law and transmit the status of its current state to the neighbors.
Then, the error between the last transmitted state and the current state for $t \in [t_{k}, t_{k+1})$ is given by
\begin{align}\label{eq406}
    e_{i}(t_{k}) = \Hat{x}_{i}(t_{k})-x_{i}(t_k)
\end{align}
where, $t_{k+1}=t_{k}+h $ and $h>0$ is the sampling period. 

Although $x_{i}(t)$ is sampled at instants decided by $h(t)$,   $\Hat{x}_{i}(t_{k})$ are updated only when the respective event condition is also satisfied. This scenario automatically guarantees Zeno-free behavior in the closed-loop system.

We define, $\Bar{\lambda}_{i}=\lambda_{max}(\lvert S_{i} \rvert^{T}\lvert S_i \rvert )$ ,   $\Bar{\lambda}=\underset{i\in\mathcal{V}}{\max}\Bar{\lambda}_{i}$, $\rho_{i}=2(\Bar{\lambda}_{i} +\Bar{\lambda})$, 
$\underline{\lambda}_{i}=\lambda_{min}\left(\left(\lvert S_{i} \rvert^{T} + \lvert S_i \rvert \right)/2\right)$,  and let $0<\sigma<1$, to propose the event-triggering function, 
\begin{align}\label{eq407}
f_{i}(e_{i}(t_{k}),\Hat{x}_{j\in N_{i}}(t_{k})) = \|S_{i}e_{i}(t_{k})\|^{2} -\frac{\sigma \underline{\lambda}^{2}_{i}}{\lvert \mathcal{N}_{i}\rvert ^{2} \rho_{i}}\|\sum_{j\in \mathcal{N}_{i}}(S_{i}\Hat{x}_{i}(t_{k})-S_{j}\Hat{x}_{j}(t_{k}))\|^{2} 
\end{align}
such that an event is triggered for the $i^{th}$ agent if 
\begin{align}\label{eq07}
 f_{i}(e_{i}(t_{k}),\Hat{x}_{j\in N_{i}}(t_{k}))\geq 0   
\end{align}
 When an event is triggered, the agent provides its current state to the controller and transmits the information to neighbors to update their control laws. Simultaneously, $e_{i}(t_{k})$ is reset to zero. The following theorem encapsulates the main result of this article.
\begin{theorem}
    Consider a matrix-scaled multi-agent network of $n$ agents modeled as in (\ref{eq02}) and interacting via an undirected and connected graph $\mathcal{G}$. Let $\beta > (1/Re(\Lambda_{1}))$, where $\Lambda_1$ is the nonzero eigenvalue of $|S|\bar{L}$ with the smallest real part. Define $\underline{\lambda}=\underset{i\in\mathcal{V}}{min}\underline{\lambda}_{i}$, and $\mathcal{N}_{max}=\underset{i\in\mathcal{V}}{\max}|\mathcal{N}_{i}|$. Then, the distributed control law 
    \begin{align}\label{eq08}
    u_{i}(t)=-\frac{\beta}{T-t}sgn(S_{i})\sum_{j\in \mathcal{N}_{i}}(S_{i}\Hat{x}_{i}(t)-S_{j}\Hat{x}_{j}(t))
    \end{align}
    together with the event-triggering protocol in (\ref{eq07}), defined for $t \in [0,T)$, drives the networked system from any initial state to the consensus space defined by (\ref{eq1}), within a user-prescribed finite-time T, provided the maximum sampling period at time $t$ is less than $(T-t)(1-e^{-\Delta})$, where $\Delta =(1-\sigma)\underline{\lambda}/(4 \beta \mathcal{N}_{max}\Bar{\lambda})$.
    Further, all control inputs $u_{i}(t)$ remain bounded for $t\in [0, T)$. 
\end{theorem}
\begin{proof}
The closed-loop dynamics of the $i^{th}$ agent can be written using (\ref{eq02}) and (\ref{eq08}) as
\begin{align}
    \Dot{x}_{i}(t)=-\frac{\beta}{T-t}sgn(S_{i})\sum_{j\in \mathcal{N}_{i}}(S_{i}\Hat{x}_{i}(t)-S_{j}\Hat{x}_{j}(t))
\end{align}
The closed-loop dynamics of the network can be expressed as
\begin{align}\label{eq0011}
    \Dot{\text{x}}(t)=-\frac{\beta}{T-t}(sgn(S)\otimes I_{d})\Bar{L}S\Hat{\text{x}}(t)
\end{align}
where, $\Hat{\text{x}}(t)=\begin{bmatrix}
   \Hat{\text{x}}_{1}^{T}(t) & \Hat{\text{x}}_{2}^{T}(t) &... &  \Hat{\text{x}}_{n}^{T}(t) 
\end{bmatrix}^{T}$ contains the latest updated states of all agents. 

Now,  applying the state transformation $\text{x}_{c}=S\text{x}(t)$
(hence, $\Hat{\text{x}}_{c}(t)=S\Hat{\text{x}}(t)$) to (\ref{eq0011}) yields
\begin{align}\label{eq412}
    \Dot{\text{x}}_{c}(t)=-\frac{\beta}{T-t}|S|\Bar{L}\Hat{\text{x}}_{c}(t)
\end{align}
The prescribed-time control of (\ref{eq412}) is converted into an equivalent asymptotic stabilization problem by using the time transformation given by
\begin{align}\label{eq413}
\tau = ln\left(\frac{T}{T-t}\right) \triangleq a(t)
\end{align}
The transformation is continuously differentiable and strictly increasing, with $a(0)=0$ and $a(T)=\infty$. Therefore, it can map the dynamics of the system, in the finite-time domain with $t\in[0,T)$ as the time parameter, into an infinite-time domain with $\tau \in [0,\infty)$ as the new time parameter. Let $\text{x}(t)= p(\tau)$ (hence, $\Hat{\text{x}}(t)= \Hat{p}(\tau)$) and $\text{x}_{c}(t)=q(\tau)$ (hence, $\Hat{\text{x}}_{c}(t)= \Hat{q}(\tau)$). Then, using the method described by (\ref{800}), the equivalent dynamics of (\ref{eq412}) in the new time frame is
\begin{align}\label{eq414}
   q'(\tau)=-\beta |S|\Bar{L}\Hat{q}(\tau)
\end{align}
Let $\epsilon_{i}(\tau)=S_{i}e_{i}(t)$ for $i=1,\dots, n$ so that transforming (\ref{eq405}) and (\ref{eq406}) yields $\epsilon_{i}(\tau) = \Hat{q}_{i}(\tau)-q_{i}(\tau) $ and $ \epsilon_{i}(\tau_{l}) = \Hat{q}_{i}(\tau_{l})-q_{i}(\tau_l), \text{for } \tau \in [\tau_{l}, \tau_{l+1}) $.
In addition, let $\tau_{l+1}=\tau_{l}+ \delta $, where $\delta$ is equivalent to its finite-time counterpart, $h=t_{k+1}-t_{k}$.
An event-triggering protocol is now developed for (\ref{eq414}) by considering the Lyapunov function in the interval $\tau \in [\tau_{l}, \tau_{l+1})$
\begin{align}\label{eq417}
    V(\tau)=(1/2)q^{T}(\tau)\Bar{L}q(\tau)
\end{align}
The time variation of $V(\tau)$ along the trajectories of (\ref{eq414}) is 
\begin{align}\label{eq418}
    V'(\tau)&=(1/2)q'^T(\tau)\Bar{L}q(\tau) + (1/2)q^T(\tau)\Bar{L}q'(\tau)\nonumber\\
    &= -\beta q^T(\tau)\Bar{L} \lvert S \rvert \Bar{L}\Hat{q}(\tau_{l})
\end{align}
From $q(\tau)=\Hat{q}(\tau_{l})-\epsilon(\tau)$, $\epsilon(\tau) = \epsilon(\tau_{l})+(\tau -\tau_{l})\beta |S|\Bar{L}\Hat{q}(\tau_{l}),$ and $\Hat{z}(\tau_{l})\triangleq \Bar{L}\Hat{q}(\tau_{l})$, where $\Hat{z}(\tau_{l})\in \mathbb{R}^{nd}$, (\ref{eq418}) simplifies to
\begin{align}
    V'(\tau) =  -\beta \Hat{z}^T(\tau_{l})\lvert S \rvert \Hat{z}(\tau_{l}) + \beta  \Hat{z}^{T}(\tau_{l})\lvert S \rvert ^{T} \Bar{L} \epsilon(\tau_{l}) 
    + \beta^{2} (\tau -\tau_{l})\Hat{z}^T(\tau_{l})\lvert S \rvert ^{T} \Bar{L}|S|\Hat{z}(\tau_{l})
    \end{align}  
We denote $V'_{1} = -\beta \Hat{z}^T(\tau_{l})\lvert S \rvert \Hat{z}(\tau_{l}) + \beta  \Hat{z}^{T}(\tau_{l})\lvert S \rvert ^{T} \Bar{L} \epsilon(\tau_{l}) $ and $ V'_{2}(\tau) = \beta^{2} (\tau -\tau_{l})\Hat{z}^T(\tau_{l})\lvert S \rvert ^{T} \Bar{L}|S|\Hat{z}(\tau_{l})$ so that 
\begin{align}\label{eq0421}
  V'(\tau) = V'_{1}(\tau)+V'_{2}(\tau) 
\end{align}
First, consider the summation form of $V'_1(\tau)$, given by
\begin{align}
V'_{1}(\tau) &= -\beta \sum\limits_{i=1}^{n} \Hat{z}_{i}^{T}(\tau_{l}) \lvert S_{i} \rvert \Hat{z}_{i}(\tau_{l})   +\beta \sum\limits_{i=1}^{n} \Hat{z}_{i}^{T}(\tau_{l}) \lvert S_{i} \rvert^{T} \left( \sum\limits_{j\in\mathcal{N}_{i}}(\epsilon_{i}(\tau_{l})-\epsilon_{j}(\tau_{l}))\right) \nonumber \\
&\leq -\beta \sum\limits_{i=1}^{n} \left( \underline{\lambda}_{i} \|\Hat{z}_{i}(\tau_{l})\|^{2}  + \lvert \mathcal{N}_{i} \rvert \Hat{z}_{i}^{T}(\tau_{l})   \lvert S_{i} \rvert^{T}\epsilon_{i}(\tau_{l})\right)  - \beta\sum\limits_{i=1}^{n} \Hat{z}_{i}^{T} (\tau_{l})\sum\limits_{j\in\mathcal{N}_{i}}\lvert S_{i}\rvert ^{T} \epsilon_{j}(\tau_{l}) \label{eq421a}
\end{align}
Applying Lemma \ref{lemma1} in (\ref{eq421a}) yields
\begin{align} \label{eq0420}
V'_{1}(\tau) &\leq -\beta \sum\limits_{i=1}^{n} \left( \underline{\lambda}_{i} \|\Hat{z}_{i}(\tau_{l})\|^{2}   + \frac{a_{i}\lvert \mathcal{N}_{i} \rvert}{2}  \|\Hat{z}_{i}(\tau_{l})\|^{2} \right) \nonumber \\ 
& \qquad + \beta \sum\limits_{i=1}^{n} (\frac{\lvert \mathcal{N}_{i} \rvert}{2a_{i}}\epsilon_{i}^{T}(\tau_{l})\lvert S_{i} \rvert^{T}\lvert S_{i} \rvert\epsilon_{i}(\tau_{l})  +  \frac{a_{i}\lvert \mathcal{N}_{i} \rvert}{2} \|\Hat{z}_{i}(\tau_{l})\|^{2})  \nonumber \\
&\qquad  + \frac{\beta}{2a_{i}}\sum\limits_{i=1}^{n}\sum\limits_{j\in\mathcal{N}_{i}}\epsilon_{j}^{T} |S_{i}||S_{i}|^{T}\epsilon_{j} \nonumber \\
& \leq -\beta \sum\limits_{i=1}^{n} \left(\underline{\lambda}_{i} -a_{i}|N_{i}| \right)\|\Hat{z}_{i}(\tau_{l})\|^{2} - \frac{\lvert \mathcal{N}_{i} \rvert \rho_{i}}{4a_{i}}\| \epsilon_{i}(\tau_{l})\|^{2}
\end{align}  
Taking $ a_{i} =\underline{\lambda}_{i}/(2\lvert\mathcal{N}_{i}\rvert)
$, one can express (\ref{eq0420}) as
\begin{align}\label{eq426}
   V_{1}'(\tau) &\leq \frac{-\beta}{2}\sum\limits_{i=1}^{n} ( \underline{\lambda}_{i} \|\Hat{z}_{i}(\tau_{l})\|^{2} - \frac{\lvert \mathcal{N}_{i}\rvert^{2}\rho_{i}}{\underline{\lambda}_{i}}  \|\epsilon_{i}(\tau_{l})\|^{2} )
\end{align}
Now, ensuring (\ref{eq427}) guarantees (\ref{eq428}), i.e., $V_{1}'(\tau)\leq 0$.
\begin{align}
    \|\epsilon_{i}(\tau_{l})\|^{2} &\leq \frac{\sigma \underline{\lambda}^{2}_{i}}{\lvert \mathcal{N}_{i}\rvert ^{2} \rho_{i}}\|\Hat{z}_{i}(\tau_{l})\|^{2} \label{eq427}\\
         V_{1}'(\tau) &\leq -\frac{\beta}{2}\sum\limits_{i=1}^{n} (1-\sigma)\underline{\lambda}_{i} \|\Hat{z}_{i}(\tau_{l})\|^{2}   \leq 0 \label{eq428}
\end{align}
We propose the desired event-triggering condition from (\ref{eq427}). Accordingly, an event is triggered for the $i^{th}$ agent when
\begin{align}\label{eq430}
    f_{i}(\epsilon_{i}(\tau_{l}), q(\tau_{l}))&=\|\epsilon_{i}(\tau_{l})\|^{2}- \frac{\sigma \underline{\lambda}^2_{i}}{\lvert \mathcal{N}_{i}\rvert ^{2} \rho_{i}}\|\Hat{z}_{i}(\tau_{l})\|^{2} \nonumber\\
     &=\|\epsilon_{i}(\tau_{l})\|^{2}- \frac{\sigma \underline{\lambda}^2_{i}}{\lvert \mathcal{N}_{i}\rvert ^{2} \rho_{i}}\|\sum_{j\in \mathcal{N}_i}(\Hat{q}_{i}(\tau_{l})-\Hat{q}_j(\tau_{l}))\|^{2} \geq 0
\end{align}
The triggering function (\ref{eq407}) is obtained by transforming (\ref{eq430}) into the finite-time domain.
Now, consider $V'_{2}(\tau)$ in (\ref{eq0421})
\begin{align}\label{eq431}
  V'_{2}(\tau) &= \beta^{2} (\tau -\tau_{l})\Hat{z}^T(\tau_{l})\lvert S \rvert ^{T} \Bar{L}|S|\Hat{z}(\tau_{l}) \nonumber \\
   & \leq \beta^{2} \delta \sum _{i=1}^{n} \Hat{z}^{T}_{i}(\tau_{l})|S_{i}|^{T}\left(\sum_{j\in \mathcal{N}_{i}}|S_{i}|\Hat{z}_{i}(\tau_{l})-|S_{j}|\Hat{z}_{j} (\tau_{l})\right) 
\end{align}
Using Lemma \ref{lemma1},  (\ref{eq431}) can be simplified as
\begin{align}\label{eq432}
   V'_{2}(\tau)  &\leq 2 \beta^{2} \delta \sum _{i=1}^{n} |\mathcal{N}_{i}|\Bar{\lambda}_{i}\| \Hat{z}_{i}(\tau_{l}) \|^{2}
\end{align}
From (\ref{eq0421}), (\ref{eq428})  and (\ref{eq432}), we have
\begin{align}\label{eq433}
   V'(\tau)&\leq -\beta^{2}\sum\limits_{i=1}^{n}\left( \frac{(1-\sigma)\underline{\lambda}_{i}}{2\beta}-2\delta|\mathcal{N}_{i}|\Bar{\lambda}_{i} \right) \|\Hat{z}_{i}(\tau_{l})\|^{2} \nonumber \\
 &\leq -\beta^{2}  \sum\limits_{i=1}^{n} \left(\frac{(1-\sigma)\underline{\lambda}}{2\beta}-2 \delta \mathcal{N}_{max}\Bar{\lambda}\right) \|\Hat{z}_{i}(\tau_{l})\|^{2} 
\end{align}
If the sampling interval is appropriately chosen to satisfy
\begin{align}\label{eq0434}
  \delta < \frac{(1-\sigma)\underline{\lambda}}{4\beta \mathcal{N}_{max}\Bar{\lambda}} \triangleq \Delta
\end{align}
then, $V'(\tau)\leq 0, \textbf{ } \forall \textbf{ }\tau \in [0,\infty)$.
Using (\ref{eq413}), the maximum sampling period in the finite-time domain becomes
\begin{align}\label{eq436}
    h(t)= (T-t)(1-e^{-\Delta})
\end{align}

Thus, the maximum sampling period is not constant but depends on the timing of previous samples. From \eqref{eq433}, \( V'(\tau) = 0 \) iff \( \hat{z}(\tau) = 0 \), i.e., when \( \Bar{L} \hat{q}(\tau) = 0 \). Now, using LaSalle's invariance principle, we obtain \( q(\tau) \rightarrow \ker(\Bar{L}) = \operatorname{im}(1_{n} \otimes I_{d}) \) as \( \tau \rightarrow \infty \). According to~\cite{timetrnasformYucelen}, the dynamics in~\eqref{eq412} over \( t \in [0, T) \) and in~\eqref{eq414} over \( \tau \in [0, \infty) \) are equivalent under the time scaling in~\eqref{eq413}. Hence, \( \text{x}_{c}(t) \rightarrow \operatorname{im}(1_{n} \otimes I_{d}) \) as \( t \rightarrow T \). 

To prove the boundedness of the control law, we introduce a scaled version of the control input given by $v(t)=Su(t)$ and show that $v(t)\rightarrow0$ as $t \rightarrow T$. Since \( S \) is invertible, the boundedness of $v(t)$ translates into the boundedness of the actual control input $u(t)$. For this purpose, we express $v(t)$ in terms of the disagreement vector and then employ time transformation to show that the scaled-control input exponentially converges to zero in the infinite-time frame. Since the time transformation is a diffeomorphism, this would establish that $v(t) \rightarrow 0$ as $t \rightarrow T$.  

Consider the proposed distributed control law in (\ref{eq08}), 
    \begin{align}
    u_{i}(t) = -\frac{\beta}{T - t} \operatorname{sgn}(S_{i}) \sum_{j \in \mathcal{N}_{i}} (S_{i}\hat{x}_{i}(t) - S_{j} \hat{x}_{j}(t)), \quad i = 1,2,\dots,n\nonumber
    \end{align}
  which can be expressed in compact form as:
    \begin{align}\label{eq2}
    u(t) &= -\frac{\beta}{T - t}(\operatorname{sgn}(S) \otimes I_d)\bar{L} S \hat{\text{x}}(t) \nonumber \\
         &= -\frac{\beta}{T - t}(\operatorname{sgn}(S) \otimes I_d)\bar{L} \hat{\text{x}}_{c}(t)
    \end{align}  
   Define \( v(t) = Su(t) \) so that (\ref{eq2}) can be expressed as
    \begin{align}\label{eq3}
    v(t) = -\frac{\beta}{T - t} |S| \bar{L} \hat{\text{x}}_{c}(t)
    \end{align}
Now, we can refer to Equation (\ref{eq412}) to infer that $\dot{\text{x}}_c(t) \equiv v(t)$. 
    Next, we define the disagreement vector for matrix-scaled networks as
    \begin{align}\label{eq04}
        \xi (t)\triangleq \text{x}_{c}(t)-(1_{n}\otimes I_{d})x^{v}
    \end{align}
    where $x^{v}$ is the virtual consensus point of a matrix-scaled network. Correspondingly, disagreement vector dynamics are given by $      \dot{\xi}(t) =\dot{\text{x}}_{c}(t)$.
    Thus, $v(t)=\dot{\xi}(t)$. We now transform the signals and corresponding dynamics in the finite-time interval $t\in [0,T)$, into the infinite-time domain $\tau \in [0, \infty)$, using the transformation $\tau = \ln(T / (T - t))$. For this purpose, we define $\omega(\tau) = v(t)$, $q (\tau)\triangleq\text{x}_{c}(t)$ and $\zeta (\tau)\triangleq\xi(t)$  so that $\zeta(\tau)=q(\tau)-(1_{n}\otimes I_{d})x^{v}$. Then, we have   
    \begin{align}
        \dot{\xi}(t) &= \frac{d\zeta}{d\tau}\frac{d\tau}{dt}=\frac{e^{\tau}}{T}\zeta '(\tau) \label{eq007} \\
        \omega (\tau) &=\dot{\xi}(t) = \frac{e^{\tau}}{T}\zeta'(\tau) \label{eq8}
   \end{align}
   Furthermore, using Equation (\ref{eq414}) and noting that $\epsilon(\tau) = \hat{q}(\tau)- q(\tau)$, $(1_{n}\otimes I_{d})\bar{L}\equiv 0$, we have
   \begin{align}
    \zeta '(\tau) &= q'(\tau) =-\beta |S| \bar{L}\hat{q}(\tau) =-\beta |S| \bar{L}(q(\tau)+\epsilon(\tau))=-\beta |S| \bar{L}  \zeta(\tau)-\beta |S| \bar{L} \epsilon(\tau) \label{eq12}
    \end{align}
    Since $\|\omega (\tau)\| \leq \frac{e^{\tau}}{T} \|\zeta '(\tau)\| $, we now proceed to solve (\ref{eq12}) analytically to obtain an upper bound for $\|\zeta '(\tau)\|$.
     
    To this end, let $P=(\sum_{i=1}^{n}|S_{i}|^{-1})^{-1}$, and $ A \triangleq|S|\Bar{L}$, whose Jordan normal form is $J$. The matrix $A$ has $d$ zero eigenvalues and $dn-d$ eigenvalues with positive real parts (see Lemma 2 in \cite{trinh_msc_conf}). Furthermore, there exists $W=\begin{bmatrix}
    w_{1} & ... & w_{dn} \end{bmatrix} \in \mathbb{C}^{dn \times dn} $ with $W_{[1:d]}=1_n\otimes I_{d}$ and $((W^{-1})^{T})_{[1:d]}=(|S^{-1}|)^{T}(1_{n}\otimes I_{d})P^{T}$ such that $A = WJW^{-1}$.  Further, the virtual consensus point, which is time-invariant, can be expressed as
    \begin{align} \label{eq014}
        x^{v}=P(1_{n}^{T}\otimes I_{d})|S^{-1}|q(\tau)
    \end{align}
       
    By definition, we have 
    \begin{align}\label{eq14}
       (((W^{-1})^{T})_{[1:d]})^{T}W_{[1:d]} &= P(1_{n}^{T}\otimes I_{d})|S^{-1}|(1_{n}\otimes I_{d}) \nonumber \\
       & = (\sum_{i=1}^{n}|S_{i}|^{-1})^{-1} (1_{n}^{T}\otimes I_{d})|S^{-1}|(1_{n}\otimes I_{d})\nonumber \\
       & =(\sum_{i=1}^{n}|S_{i}|^{-1})^{-1}  (\sum_{i=1}^{n}|S_{i}|^{-1}) \nonumber \\ 
       &= I_{d}
    \end{align}
   The analytical solution of (\ref{eq12}) for $\tau \in [\tau_{l},\tau_{l+1})$ with $\zeta(\tau_{l})=\zeta_{l}$ is 
   \begin{align} \label{eq015}
       \zeta(\tau) &=  e^{-\beta A (\tau-{\tau_{l})}}\zeta_{l}-\beta \int_{\tau_{l}}^{\tau}  e^{-\beta A(\tau-s)}A\epsilon(s)ds  
\end{align}
Now, we express the matrix exponential $e^{-\beta A \tau}$ as
       \begin{align}  
        e^{-\beta A \tau}&= We^{-\beta J\tau}W^{-1} \nonumber \\
        &= W_{[1:d]}(((W^{-1})^{T})_{[1:d]})^{T} +WD_{\tau}W^{-1}
   \end{align}
where $D_{\tau}\triangleq diag\left(\begin{bmatrix}
  zeros(1,d)  & e^{-\beta J_{1}\tau}  & ... & e^{-\beta J_{p}\tau}
\end{bmatrix}\right)$ and $J_{1}$, \dots,  $J_{p}$ are the Jordan blocks corresponding to the nonzero eigenvalues $\Lambda_1$, $\Lambda_2$, \dots, $\Lambda_p$, $p\leq dn-d$ of the matrix $A$. 
Since
\begin{align}\label{eq17}
    W_{[1:d]}(((W^{-1})^{T})_{[1:d]})^{T}\zeta(\tau) &= (1_n\otimes I_{d})P(1^{T}_n\otimes I_{d})|S|^{-1}\zeta(\tau) \nonumber \\
    & = (1_n\otimes I_{d})PP^{-1}x^{v}-(1_{n}\otimes I_{d})x^{v} \nonumber \\ &\equiv 0
\end{align}
and
\begin{align}
     W_{[1:d]}(((W^{-1})^{T})_{[1:d]})^{T}A \epsilon(\tau) & = (1_n\otimes I_{d})P(1^{T}_n\otimes I_{d})|S|^{-1}|S|\bar{L} \epsilon(\tau) \nonumber \\
   & = (1_n\otimes I_{d})P(1^{T}_n\otimes I_{d})\bar{L} \epsilon(\tau) \nonumber \\ &\equiv 0
\end{align}
Equation (\ref{eq015}) simplifies to
\begin{align} \label{eq15}
       \zeta(\tau)=WD_{(\tau - \tau_{l})}W^{-1}\zeta_{l} - \beta \int_{\tau_{l}}^{\tau}  WD_{(\tau -s)}W^{-1}A\epsilon(s)ds 
   \end{align}
where $D_{(\tau - \tau_{l})}= \begin{bmatrix}
  zeros(1,d)  & e^{-\beta J_{1}(\tau -\tau_{l})}  & ... & e^{-\beta J_{p}(\tau -\tau_{l})}\end{bmatrix}$. Let $c_{1}=\|W\| \| W^{-1}\| $ and $Re(\Lambda_{1})< \text{ } ... \text{ } < Re(\Lambda_{p})$. Then, 
\begin{align}
   \|\zeta(\tau)\| &\leq  c_1e^{-\beta Re(\Lambda_{1})(\tau-\tau_{l}}) \|\zeta_{l}\| +\beta c_{1}\| A\| \int_{\tau_{l}}^{\tau}  e^{-\beta Re(\Lambda_{1})(\tau-s)}\|\epsilon(s)\| ds \nonumber \\
   & \leq c_1e^{-\beta Re(\Lambda_{1})(\tau-\tau_{l})} \|\zeta_{l}\| +\beta c_{1}\| A\|\sqrt{n}\bar{\epsilon}_{l}  \int_{\tau_{l}}^{\tau}  e^{-\beta Re(\Lambda_{1})(\tau-s)} ds \label{eq0016} 
\end{align}
where $\|\epsilon_{i}(\tau_{l})\|\leq \sqrt{(\sigma \underline{\lambda}^2_{i})/(\lvert \mathcal{N}_{i}\rvert ^{2} \rho_{i})}\|\Hat{z}_{i}(\tau_{l})\|\triangleq \bar{\epsilon}^{l}_{i}$, and $\bar{\epsilon}_{l} = \displaystyle\max_{1 \leq i \leq n} \bar{\epsilon}^{l}_{i}$, so that $\| \epsilon(\tau)\|\leq \sqrt{n}\bar{\epsilon}_{l}$. Evaluating the integral in (\ref{eq0016}) and simplifying yields
\begin{align}\label{eq0017}
     \|\zeta(\tau) \| \leq c_{2}\Bar{\epsilon}_{l} + e^{-\beta Re(\Lambda_{1})(\tau-\tau_{l})} \left(c_{1}\|\zeta_{l}\| -  c_{2}\bar{\epsilon}_{l}\right) 
\end{align}
where $c_2=  (c_{1} \| A\|\sqrt{n})/Re(\Lambda_{1})$. Inspection of (\ref{eq0017}) reveals that during any sampling interval, the norm of the disagreement vector $\| \zeta(\tau)\|$ (hence $\|z(\tau)\|=\|\bar{L}q(\tau)\|$)  decays exponentially at a rate of $\beta Re(\Lambda_{1})$. Consequently, by definition, $\bar{\epsilon}_{l}$ also decreases over successive sampling instants in such a manner that $\bar{\epsilon}_{l}$ over $[0,\infty)$ can be upper bounded by a decaying exponential of the same rate $\beta Re(\Lambda_{1})$. Therefore, it is straight forward to conclude that there exists $k_1>0$ such that $\|\zeta(\tau) \| \leq k_1e^{-\beta Re(\Lambda_{1})\tau}\textbf{ }\forall \textbf{ } \tau \in [0, \infty)$. Similarly, (\ref{eq12}) implies that there exists $k_2 >0$ such that $\|\zeta'(\tau)\|$ can be upper bounded as
   \begin{align}\label{eq016}
   \|\zeta'(\tau)\| \leq k_2 e^{-\beta Re(\Lambda_{1})\tau} 
\end{align} 
Finally, an upper bound for $\|\omega(\tau)\|$ can be derived using (\ref{eq8}) and (\ref{eq016}) as
   \begin{align}
   \|\omega(\tau)\| &\leq  \frac{k_2}{T} e^{-\beta Re(\Lambda_{1})\tau}e^{\tau} \nonumber \\
   &\leq  \frac{k_2}{T}e^{-(\beta Re(\Lambda_{1})-1)\tau} 
\end{align} 
     If the parameter $\beta$ is selected such that $\beta Re(\Lambda_{1})>1$, then $\omega(\tau)\rightarrow 0$ as $\tau \rightarrow \infty$. Since $\omega(\tau)=v(t)$, and $t\rightarrow T$ as $\tau \rightarrow \infty $, it is immediate to conclude that $v(t)$ (hence $u(t)$) $\rightarrow 0$ as $t\rightarrow T$.   The proof is complete.
\end{proof}

The maximum sampling period given by (\ref{eq436}) linearly decreases to zero as time approaches $T$ and therefore cannot be implemented as such. 
In addition, although the overall control in (\ref{eq08}) is bounded, it contains a time-varying gain $\beta/(T-t)$ that grows to infinity as $t\rightarrow T$. This can lead to round-off errors or overflow in digital controllers and may cause instability in the computed control signals. These challenges will be addressed in future work by developing a practical prescribed-time event-triggered controller.
\section{Conclusion}
A prescribed-time event-triggered control strategy is proposed for matrix-scaled networks to achieve convergence within a user-defined time bound with reduced agent interactions. The controller implements a state-dependent triggering threshold and predefined sampling period to exclude Zeno phenomenon in the closed-loop response. The method extends the applicability of matrix-scaled consensus to complex control problems such as clustering, cooperative engagement, and dynamic formation switching. 

\bibliographystyle{IEEEtran}
\input{main.bbl}

\end{document}

%% file: main.bbl

%% file: main.bbl
\begin{thebibliography}{10}
\providecommand{\url}[1]{#1}
\csname url@samestyle\endcsname
\providecommand{\newblock}{\relax}
\providecommand{\bibinfo}[2]{#2}
\providecommand{\BIBentrySTDinterwordspacing}{\spaceskip=0pt\relax}
\providecommand{\BIBentryALTinterwordstretchfactor}{4}
\providecommand{\BIBentryALTinterwordspacing}{\spaceskip=\fontdimen2\font plus
\BIBentryALTinterwordstretchfactor\fontdimen3\font minus \fontdimen4\font\relax}
\providecommand{\BIBforeignlanguage}[2]{{%
\expandafter\ifx\csname l@#1\endcsname\relax
\typeout{** WARNING: IEEEtran.bst: No hyphenation pattern has been}%
\typeout{** loaded for the language `#1'. Using the pattern for}%
\typeout{** the default language instead.}%
\else
\language=\csname l@#1\endcsname
\fi
#2}}
\providecommand{\BIBdecl}{\relax}
\BIBdecl

\bibitem{survey_general}
Y.~Cao, W.~Yu, W.~Ren, and G.~Chen, ``An overview of recent progress in the study of distributed multi-agent coordination,'' \emph{IEEE Trans. Ind. Inform.}, vol.~9, no.~1, pp. 427--438, 2013.

\bibitem{secure}
M.~Fabris and D.~Zelazo, ``A robustness analysis to structured channel tampering over secure-by-design consensus networks,'' \emph{IEEE Control Syst. Lett.}, vol.~7, pp. 2011--2016, 2023.

\bibitem{zhao2014similitude}
L.~Zhao, ``Similitude design method for motion reconstruction of space cooperative vehicles,'' \emph{J. Aastronaut.}, vol.~35, no.~7, p. 802, 2014.

\bibitem{haddad2010nonnegative}
W.~M. Haddad, V.~Chellaboina, and Q.~Hui, \emph{Nonnegative and compartmental dynamical systems}.\hskip 1em plus 0.5em minus 0.4em\relax Princeton University Press, 2010.

\bibitem{ROY2015259}
S.~Roy, ``Scaled consensus,'' \emph{Automatica}, vol.~51, pp. 259--262, 2015.

\bibitem{liu2025}
Y.~Liu, Y.~Wang, H.~Zhang, and S.~Yang, ``Scaled consensus of multiagent systems with unknown input delay and disturbance,'' \emph{IEEE Trans. on Autom. Sci. Eng.}, vol.~22, pp. 18\,431--18\,442, 2025.

\bibitem{mo2022hierarchical}
S.~Mo, W.-H. Chen, and X.~Lu, ``Hierarchical hybrid control for scaled consensus and its application to secondary control for dc microgrid,'' \emph{IEEE Trans. Cybern.}, vol.~53, no.~7, pp. 4446--4458, 2022.

\bibitem{trinh_msc_conf}
M.~H. Trinh, D.~Van~Vu, Q.~V. Tran, and H.-S. Ahn, ``Matrix-scaled consensus,'' in \emph{Proc. IEEE Conf. Decis. Control}, 2022, pp. 346--351.

\bibitem{trinh_msc}
M.~H. Trinh, H.~H. Vu, N.-M. Le-Phan, and Q.~N. Nguyen, ``Matrix-scaled consensus over undirected networks,'' \emph{IEEE Trans. Netw. Sci. Eng.}, vol.~11, no.~6, pp. 6678--6691, 2024.

\bibitem{zhang2024containment}
X.~Zhang, L.~Pan, H.~Shao, and D.~Li, ``Containment control of matrix-scaled multi-agent networks,'' in \emph{Proc. 2024 IEEE Int. Conf. Unmanned Syst.}\hskip 1em plus 0.5em minus 0.4em\relax IEEE, 2024, pp. 1660--1665.

\bibitem{MSCuncertainnetworks}
B.~Cheng and G.~Li, ``Matrix-scaled consensus of uncertain networked systems,'' in \emph{Proc. Chin. Control Conf. (CCC)}, 2023, pp. 5981--5985.

\bibitem{secondorderdirected}
S.~Chen, K.~Tian, and J.~Mei, ``Matrix-scaled consensus for second-order uncertain multi-agent systems under a directed graph,'' in \emph{Proc. Chin. Control Conf. (CCC)}, 2023, pp. 6111--6116.

\bibitem{MSCWeightedNetworksWithStateConstraintsstateconstraints}
Y.~Shang, ``Matrix-scaled consensus on weighted networks with state constraints,'' \emph{IEEE Syst. J.}, vol.~17, no.~4, pp. 6472--6479, 2023.

\bibitem{finitetimehomogenous}
X.~Wang, S.~Li, and P.~Shi, ``Distributed finite-time containment control for double-integrator multiagent systems,'' \emph{IEEE Trans. Cybern.}, vol.~44, no.~9, pp. 1518--1528, 2014.

\bibitem{fixedtimeoverview}
Z.~Zuo, Q.-L. Han, B.~Ning, X.~Ge, and X.-M. Zhang, ``An overview of recent advances in fixed-time cooperative control of multiagent systems,'' \emph{IEEE Trans. Ind. Inform.}, vol.~14, no.~6, pp. 2322--2334, 2018.

\bibitem{PTCfirst}
Y.~Song, Y.~Wang, J.~Holloway, and M.~Krstic, ``Time-varying feedback for regulation of normal-form nonlinear systems in prescribed finite time,'' \emph{Automatica}, vol.~83, pp. 243--251, 2017.

\bibitem{PTcontrainmentcontrol}
Y.~Wang, Y.~Song, D.~J. Hill, and M.~Krstic, ``Prescribed-time consensus and containment control of networked multiagent systems,'' \emph{IEEE Trans. Cybern.}, vol.~49, no.~4, pp. 1138--1147, 2019.

\bibitem{PTcooperativeengagement}
T.~Yucelen, Z.~Kan, and E.~Pasiliao, ``Finite-time cooperative engagement,'' \emph{IEEE Trans. Autom. Control}, vol.~64, no.~8, pp. 3521--3526, 2019.

\bibitem{ptcreviewlewis}
Y.~Song, H.~Ye, and F.~L. Lewis, ``Prescribed-time control and its latest developments,'' \emph{IEEE Trans. Syst., Man, Cybern. Syst.}, vol.~53, no.~7, pp. 4102--4116, 2023.

\bibitem{su2025prescribed}
J.~Su and Y.~Song, ``Prescribed-time control with bounded feedback gain: A nonscaling and structural adaptation-based approach,'' \emph{IEEE Trans. Syst., Man, Cybern.: Syst.}, 2025.

\bibitem{decentralized2025}
H.~Ye, C.~Wen, and Y.~Song, ``Decentralized and distributed control of large-scale interconnected multiagent systems in prescribed time,'' \emph{IEEE Trans. Autom. Control}, vol.~70, no.~2, pp. 1115--1130, 2025.

\bibitem{eventtrigger}
G.~S. Seyboth, D.~V. Dimarogonas, and K.~H. Johansson, ``Event-based broadcasting for multi-agent average consensus,'' \emph{Automatica}, vol.~49, no.~1, pp. 245--252, 2013.

\bibitem{eventtriggerold}
M.~Guinaldo, D.~V. Dimarogonas, K.~H. Johansson, J.~Sanchez, and S.~Dormido, ``Distributed event-based control strategies for interconnected linear systems,'' \emph{ET Control Theory Appl.}, vol.~7, no.~6, pp. 877--886, 2013.

\bibitem{timetrnasformYucelen}
D.~Kurtoglu and T.~Yucelen, ``A time transformation approach to finite-time distributed control with reduced information exchange,'' \emph{IEEE Control Syst. Lett.}, vol.~6, pp. 458--463, 2022.

\bibitem{ptceventcluster}
F.~Zuo, Z.~Yu, and H.~Jiang, ``Prescribed-time practical cluster consensus for multi-agent systems with event-triggered control,'' in \emph{Proc. 43rd Chin. Control Conf. (CCC)}, 2024, pp. 5852--5859.

\bibitem{ptcdynamicevent}
S.~J. Yoo and B.~S. Park, ``Dynamic event-triggered prescribed-time consensus tracking of nonlinear time-delay multiagent systems by output feedback,'' \emph{Fractal Fract.}, vol.~8, no.~9, 2024.

\bibitem{eventmatrixweighted}
L.~Pan, C.~Wang, H.~Shao, D.~Li, and Y.~Xi, ``Event-triggered consensus on matrix-weighted networks,'' in \emph{Proc. China Autom. Congr. (CAC)}, 2021, pp. 7521--7525.

\end{thebibliography}
